\documentclass[pra,twocolumn,superscriptaddress,showpacs,amsmath,amssymb]{revtex4-2}

\usepackage{graphicx}% Include figure files
\usepackage{dcolumn}% Align table columns on decimal point
\usepackage{bm}% bold math
\usepackage{epsfig}
\usepackage{amsmath}
\usepackage{amssymb}
\usepackage{color}
\usepackage{bbm}
\usepackage{CJK}
\usepackage[colorlinks=true,citecolor=blue,linkcolor=blue,urlcolor=blue]{hyperref}
\begin{document}

\title{Cat-like non-Gaussian entanglement in magnon systems}
\author{Zeyu Zhang}
\affiliation{Zhejiang Key Laboratory of Quantum State Control and Optical Field Manipulation,Department of Physics, Zhejiang Sci-Tech University, 310018 Hangzhou, China}
\author{Clemens Gneiting}
\affiliation{Theoretical Quantum Physics Laboratory, Cluster for Pioneering Research, RIKEN, Wakoshi, Saitama 351-0198, Japan}
\affiliation{Center for Quantum Computing, RIKEN, Wakoshi, Saitama 351-0198, Japan}
\author{Zheng-Yang Zhou}
\altaffiliation[zheng-yang.zhou@zstu.edu.cn]{}
\affiliation{Zhejiang Key Laboratory of Quantum State Control and Optical Field Manipulation,Department of Physics, Zhejiang Sci-Tech University, 310018 Hangzhou, China}
\affiliation{Theoretical Quantum Physics Laboratory, Cluster for Pioneering Research, RIKEN, Wakoshi, Saitama 351-0198, Japan}
\author{Ai-Xi Chen}
\altaffiliation[aixichen@zstu.edu.cn]{}
\affiliation{Zhejiang Key Laboratory of Quantum State Control and Optical Field Manipulation,Department of Physics, Zhejiang Sci-Tech University, 310018 Hangzhou, China}
\date{\today}

%%%%
\begin{abstract}
%%%%
Magnons can serve as a bridge between spin, phonon, and photon systems, which renders them suitable for constructing hybrid systems. An important application of such hybrid systems is generating entanglement between different platforms. As magnons can support a broad variety of states, e.g., Fock states, squeezed states, or coherent states, and hybrid states can be produced with cavities or spins,  many different kinds of entangled states in magnon systems. In this work, we consider the entanglement of cat-like states, which can be generated in magnon systems with parametric pumps beyond the parametric stable intensities. However,  estimating the entanglement in such states is challenging due to their multi-photon and non-Gaussian properties. Here, we apply a modular variable-based projection, which maps the cat-like states to spin states, preserving the encoded information. After the projection, a Bell's inequality is employed to detect the entanglement in the effective spin states. Our numerical analysis provides the conditions for generating cat-like entanglement in magnon systems and can be conveniently extended to other entangled states that may be formed by magnon and spin systems.
\end{abstract}
\maketitle

%\pacs{
%05.30.Rt, 	% Quantum phase transitions
%42.50.Ct 	% Quantum description of interaction of light and matter; related experiments
%75.10.Jm 	% Quantized spin models, including quantum spin frustration
%        }

% ---------------------------------------------------------------------------
%
%-------------------------------------------------------------------------
%

%%%%%%%%%%%%%%%%

\section{Introduction}\label{sec:intro}
In recent years, light-matter interaction has attracted much interest~\cite{intro1,intro2,intro3}. An important example are magnon-based hybrid quantum systems~\cite{intro4,intro5,intro6,intro7,intro8,intro9,intro10,intro11,intro12,intro13,intro14}, which utilize yttrium iron garnet (YIG)  spheres~\cite{YIG1,YIG2,YIG3} that combine the advantages of a high rotational magnon effect, low saturation magnetisation intensity, small resonance line width, high resistivity, and low dielectric loss. For these reasons,  cavity-magnon systems formed by YIG spheres and microwave cavities have become an important tool for studying various quantum phenomena~\cite{intro_1,intro_2,intro_3}. Due to the strong interaction between the collective spin states in YIG crystals and the photons in the cavities, magnons play a  crucial role in the construction of quantum networks~\cite{magnon1,network1,network2,network3}. In addition, magnon-based hybrid quantum systems can be used to detect different kinds of signals, which provides a good platform for quantum sensors~\cite{magnon2,magnon3,magnon4,magnon5}. Such hybrid systems formed by magnons with photons or phonons also play a crucial role in exploring quantum information~\cite{magnon6,magnon7,magnon8}.

As a typical quantum effect without classical counterpart, quantum entanglement has many important applications in  fields such as quantum computing~\cite{entangle1,entangle2,entangle3,entangle4}, quantum metrology~\cite{metrology1,metrology2,metrology3,metrology4}, and quantum sensors~\cite{sensor1,sensor2,sensor3,sensor4}. In addition, due to the high communication security based on quantum entanglement, it is of great significance for the development of information security~\cite{information1,information2,information3,information4}. Recently, great progress has been achieved in realizing entangled magnon systems in the Gaussian (or parametric stable) regime~\cite{Gaussian1,Gaussian2,Gaussian3} or photon-number non-Gaussian regime~\cite{non-Gaussian1,non-Gaussian2,non-Gaussian3}. On the other hand non-Gaussian entanglement of magnons in the parametric unstable regime, which is important in  fields like quantum communication~\cite{communication1,communication2,communication3,communication4,communication5}, quantum sensing~\cite{sensing1,sensing2,sensing3,sensing4,sensing5}, or quantum computing~\cite{entanglement1,entanglement2,entanglement3}, is 
less explored. 

Cat-like states are a  characteristic class of non-Gaussian states in the parametric unstable regime, which can be prepared on various experimental platforms~\cite{catstate1,catstate2,catstate3}. However, the entanglement of cat-like states cannot be detected by the covariance matrix, which is in contrast to Gaussian states. An alternative viable approach employs a Bell inequality~\cite{bell1,bell2,bell3,bell4} as the cat-like states have typical Einstein-Podolsky-Rosen form~\cite{EPR1,EPR2,EPR3,EPR4,EPR5,EPR6}. Although the non-orthogonality of the coherent basis states prevents a direct application of Bell inequality, a modular variable-based projection~\cite{modular1,modular2,modular3,modular4,modular5,modularvariable1,modularvariable2} can map these states onto qubit states and preserve most information encoded in the cat-like states.

In this work, we study the conditions for generating non-Gaussian entanglement with magnon systems in the parametric unstable regime. Our model is based on  parametrically pumped magnon systems, which are typically used in studying Gaussian entanglement. For a single-mode magnon system, we numerically show that a transient cat-like state can be generated in the parametric unstable regime with the help of the Kerr effect. This result for single-mode magnon systems is then extended to  two-mode magnon systems. In addition, collective loss is introduced to generate entanglement and stabilize the state. Compared to the previously studied measurement-feedback approach~\cite{magnon8}, this method can generate steady cat states and entanglement between different magnon modes. To characterize the entanglement, we map the states to effective spin states with modular variables. In the effective spin space, a Bell inequality is applied to detect the entanglement in the system. We also explore the conditions for generating entanglement in the presence of detrimental effects like cross-talk coupling or local single-photon loss.

In Sec.~\ref{1} of this paper, a single-mode magnon system is discussed along with the corresponding Wigner function, and the parameter effects are analyzed. Based on this, a two-mode magnon system is constructed, followed by a fidelity analysis, after which the concepts of Bell's inequality and modular variables are introduced. Sec.~\ref{2} clarifies under which conditions the two-mode magnon system is in an entangled state or not, after which the coupling term and single-photon dissipation are introduced and their effects on the system are analysed. In  Sec.~\ref{3}, we draw our conclusions.\\
\section{Theoretical model}\label{1}
\label{method}
\subsection{Non-Gaussian states in  single magnons}\label{Cat state}
\begin{figure}[t]
\centering
\includegraphics[width=0.5\textwidth]{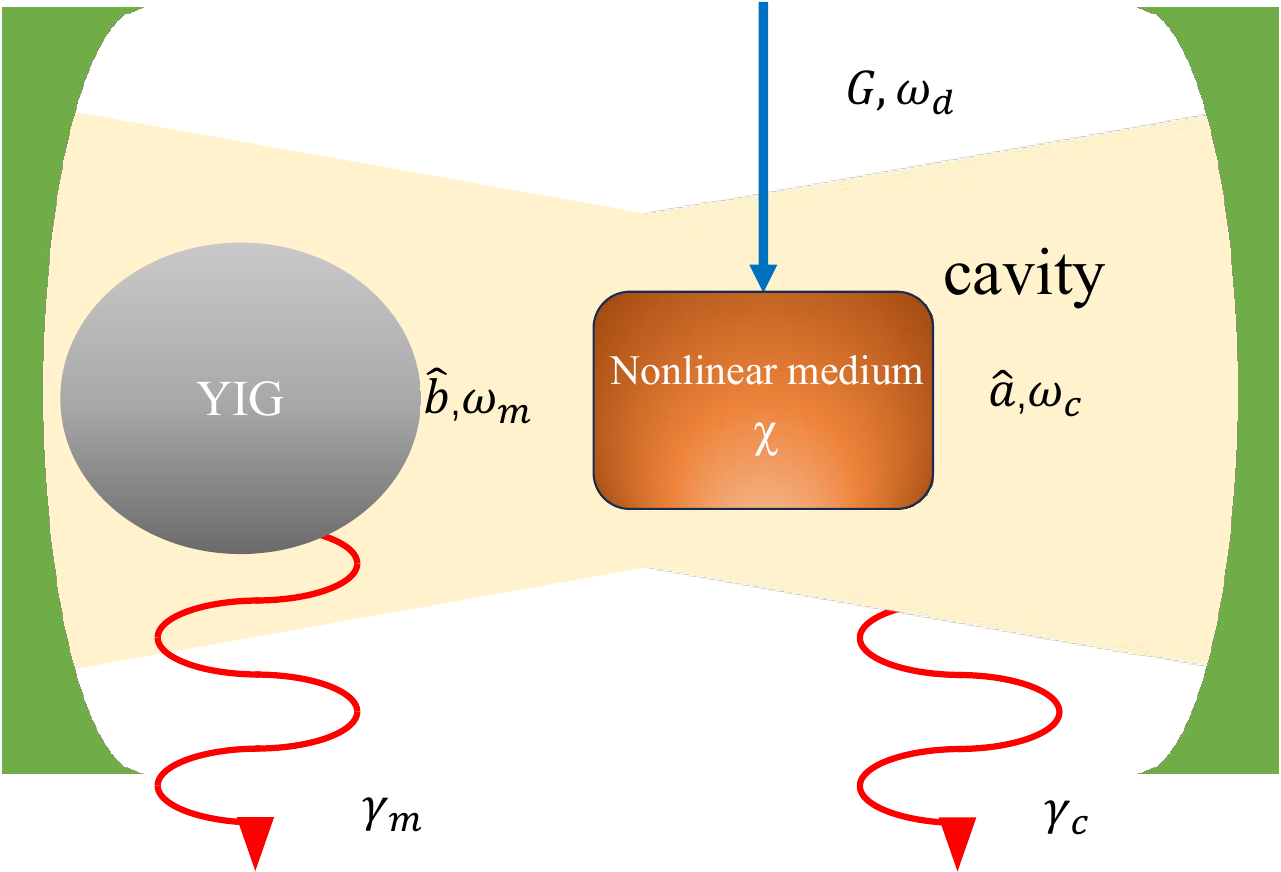}
\caption{Illustration of a single-mode cavity magnon system with a yttrium iron garnet (YIG)  sphere and a parametrically driven microwave cavity. $\hat{a}^{\dagger}$ ($\hat{a}$) denotes the creation (annihilation) operator of the microwave cavity, and $\hat{b}^{\dagger}$ ($\hat{b}$) denotes the corresponding operator for the YIG sphere. The frequencies of the cavity and the magnon are expressed by $\omega_{\rm c}$ and $\omega_{\rm m}$, respectively. The loss rates of cavity  and  magnon are denoted by $\gamma_{\rm c}$ and $\gamma_{\rm m}$, respectively. The blue arrow represents the light field driven by the medium, with  frequency  $\omega_{\rm d}$  and  amplitude  $G$. With a strong cavity loss, the cavity can be adiabatically eliminated to result in  an effective nonlinear pump on the magnon.}
\label{singe_mode}
\end{figure}

We consider a single-mode magnon coupled to a parametrically driven microwave cavity~\cite{intro12,cat2,cat3}, as illustrated in Fig.~\ref{singe_mode}. The coupled magnon and the cavity can be described by the Hamiltonian $H_{\rm s}$ (setting $\hbar$ =1):
\begin{equation}\label{cavitymagnon}
    \begin{split}
        H_{\rm s}=\omega_{\rm c}\hat{a}^{\dagger}\hat{a}+\omega_{\rm m}\hat{b}^{\dagger}\hat{b}+(g\hat{a} ^{\dagger}\hat{b}+g^*\hat{a}\hat{b}^{\dagger}),      
    \end{split}
\end{equation}
where $\hat{a}$ and $\hat{b}$ are the annihilation operators of the cavity and the magnon, respectively, $\omega_{\rm c}$ and $\omega_{\rm m}$ are the corresponding frequencies, and the third term represents the coupling term between the cavity and the magnon with the coupling strength $g$.

Due to the anisotropy of the YIG ball, a Kerr term arises~\cite{kerr},
\begin{equation}\label{kerrterms}
    \begin{split}
    H_{\rm kerr}= \frac{K}{2} \hat{b}^{\dagger}\hat{b}^{\dagger}\hat{b}\hat{b},
   \end{split}
\end{equation}  
with $K$  the  Kerr coefficient. 

To achieve  nonlinear pumping of the magnon, we  introduce a two-photon pump in the cavity~\cite{cat3}:\\
\begin{equation}\label{two-photonpump}
    \begin{split}
        H_{\rm d}=\frac{G^{*}}{2}\hat{a}^{\dagger} \hat{a}^{\dagger} e^{-i\omega_{\rm d}t }+\frac{G}{2}\hat{a}\hat{a}e^{\omega_{\rm d}t },         
    \end{split}
\end{equation}
with the two-photon pump strength $G$ and the frequency of the two-photon pump $\omega_{\rm d}\approx2\omega_{\rm m}$.

When the detuning of the cavity is large $[(\omega_{\rm c}-\omega_{\rm m})\gg g]$, we can adiabatically eliminate~\cite{adiabatic-elimination1,adiabatic-elimination2,single-photon1} the cavity mode (see Appendix \ref{adiabatic elimination}) and obtain a single-mode magnon system with nonlinear pump:\\
\begin{equation}
    \begin{split}\label{singlenonlinearmagnon}
     H =\Delta\hat{b}^{\dagger} \hat{b}+S^*\hat{b}^{\dagger} \hat{b}^{\dagger}+S\hat{b} \hat{b} +\frac{K}{2}\hat{b}^{\dagger} \hat{b}^{\dagger}\hat{b}\hat{b}, 
      \end{split}
\end{equation}
with the effective detuning $\Delta=\Delta _{\rm m}-\left |g\right | ^{2}\Delta _{\rm c}/ (\Delta _{\rm c}^{2}-\left |G \right | ^{2})+\Delta_{\rm other}$, and the effective two-photon pump rate $S = \left |g\right | ^{2} G / 2(\Delta _{\rm c}^{2}-\left |G\right | ^{2})  $. Note that the second term in the detuning $\Delta$ is caused by the effective nonlinear pumping, and $\Delta_{\rm other}$ describes the detuning caused by other cavity modes (see also Appendix~\ref{adiabatic elimination}). As these effects can be compensated by the pump frequency $\omega_{\rm d}$, we will not discuss them in this work.

\begin{figure*}[t]
  \centering
  \includegraphics[width=\textwidth]{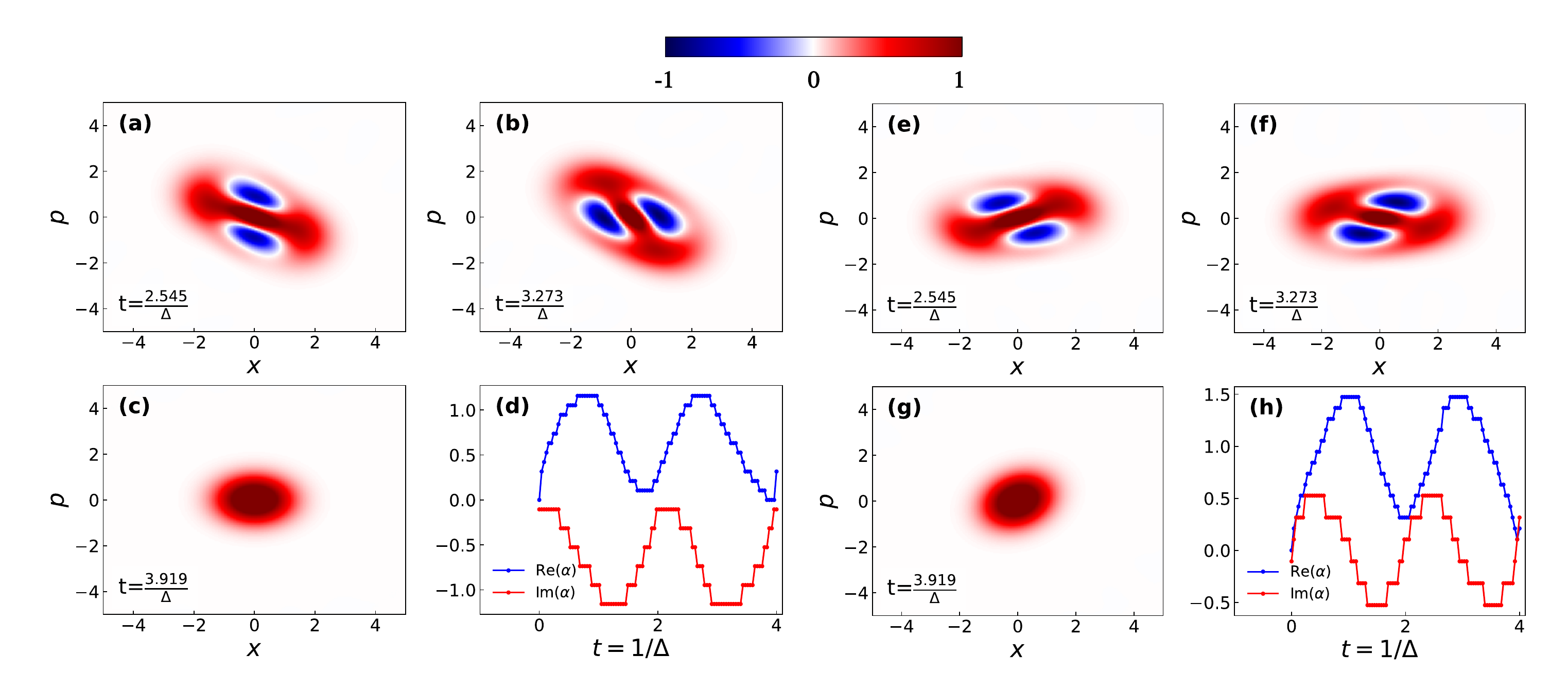}
  \caption{Numerical verification of single-mode cat state generation with different nonliner pump phases. The parameters are expressed as the ratio to $\Delta$. The Kerr intensity is set to be $K/\Delta=1.2$, and the Fock space truncation is set to be $N=15$. The nonlinear pump intensities are set to be $S/\Delta=1.2i$ for (a)--(d), and $S/\Delta=-1.2$ for (e)--(h). The initial state of the system  is $\left | \psi _{\rm 0} \right \rangle =\left | 0 \right \rangle $. (a)--(c) and (e)--(g) depict the Wigner function of the system at different times with two different nonliner pump phases, respectively. Transient cat states are prepared at different time points in both cases, as shown in (a), (b), (e) and (f). The system periodically returns to the vacuum state as shown in (c) and (g). (d) and (h) The optimal complex amplitude $\alpha$ corresponding to the cat state $1/\sqrt{N}(|\alpha\rangle+|-\alpha\rangle)$ with the highest fidelity with respect to the system state at different times. The results indicate the oscillation of the system state among the states shown in (a)-- (c) and (e)--(g).}
  \label{bestlastzzy}
\end{figure*}

Such a system can, due to the nonlinear terms, generate cat states, which are highly non-Gaussian~\cite{kerrcat1}\cite{kerrcat2}. To confirm this, we use Qutip~\cite{qutip} and Wigner functions~\cite{Wigner} to evaluate the resulting states  numerically. We set $\Delta=1$ and express other quantities in  units of $\Delta$, specifically, $S/\Delta=1.2i$ and $K/(2\Delta)=0.6$. The Wigner function plots in Fig.~\ref{bestlastzzy}(a)--(c) indicate presence of a cat state in the system. This demonstrates that the system described by Eq.~(\ref{singlenonlinearmagnon}) can exhibit strong non-Gaussian effects. Note that photon subtraction and measurements can generate similar effects in linear systems~\cite{magnon8}.\\
We emphasize that the cat state generated in our system is not a steady cat state, as no loss is introduced to stabilize the state. The results in Fig.~\ref{bestlastzzy} also confirm this. The system state oscillates between cat states with different phases and the vacuum state. To illustrate this, we optimize at all times the fidelity of the system state with respect to a cat state with  variable amplitude $\alpha$,
\begin{eqnarray}
|\psi\rangle=\frac{1}{\sqrt{2+\epsilon_{\rm s}}}(|\alpha\rangle+|-\alpha\rangle),\nonumber
\end{eqnarray}

and depict the optimal amplitudes in Fig.~\ref{bestlastzzy}(d) and (h). Note that $|\alpha\rangle$ is the coherent state of a magnon with $\hat{b}\left | \alpha  \right \rangle=\alpha\left | \alpha  \right \rangle$. From the optimal amplitudes, we can infer the waiting times for generating desired cat states , along with the corresponding sizes of the cat states. Here, the Wigner function is defined with respect to dimensionless position and momentum:
\begin{eqnarray}\label{ndxp}
\hat{x} &=& \frac{1}{\sqrt{2}} (\hat{b} + \hat{b}^{\dagger}), \nonumber\\
\hat{p} &=& -\frac{i}{\sqrt{2}} (\hat{b} - \hat{b}^{\dagger}).
\end{eqnarray}

In Fig.~\ref{bestlastzzy}(e)--(h) we consider a different choice of the nonlinear pumping phase $S/\Delta=-1.2$ and keep other parameters unchanged. While the time evolution in Fig.~\ref{bestlastzzy}(h) differs from the one in Fig.~\ref{bestlastzzy}(d), cat states can still be generated. As both phases choices can generate cat states, we assume in the following a real $S$ to simplify the expressions.
\subsection{Non-Gaussian entanglement in two dissipatively coupled magnon systems}
\begin{figure}[t]
%\captionsetup{justification=raggedright}
\centering
\includegraphics[width=0.5\textwidth]{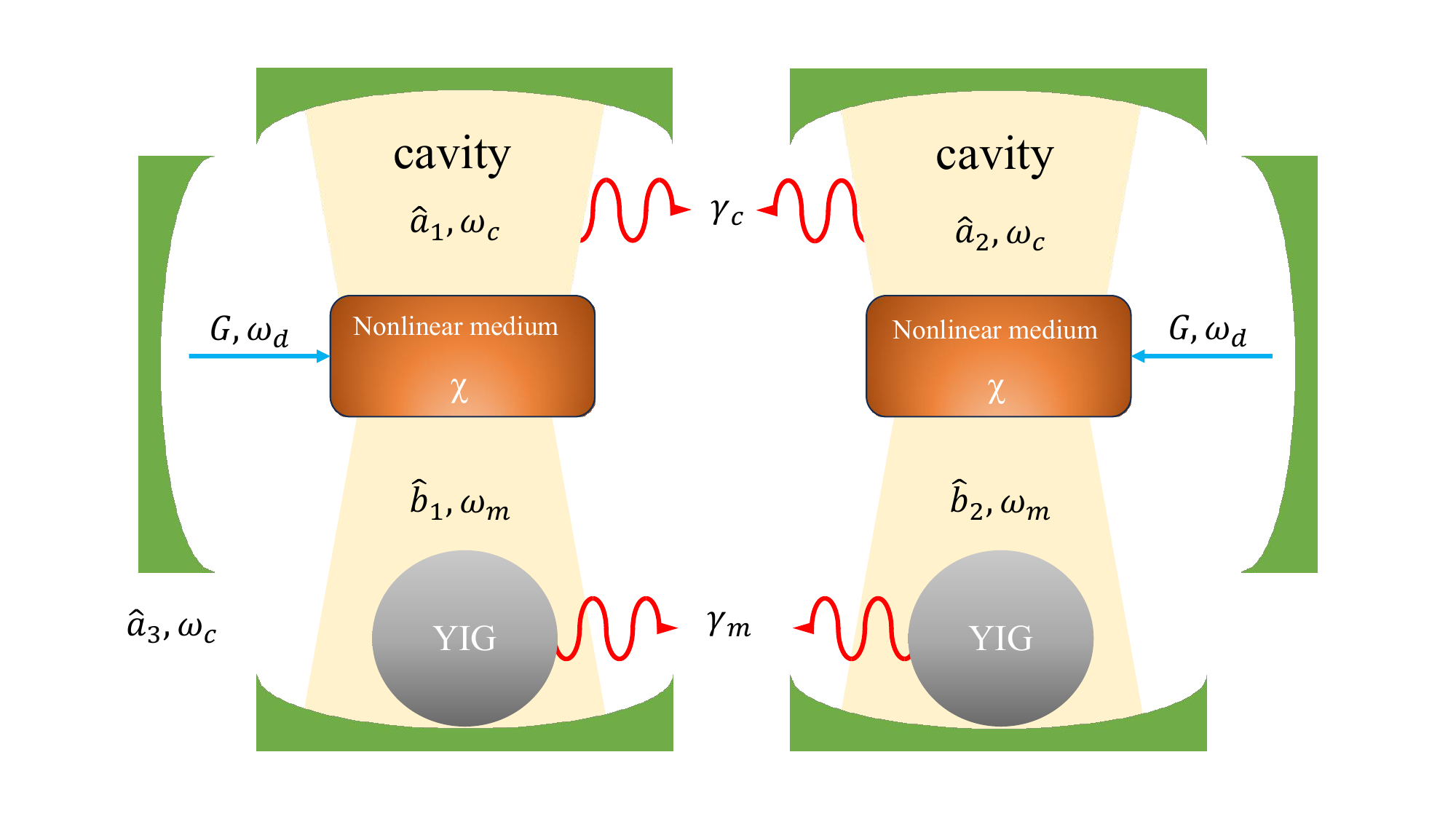}\caption{Illustration of the dissipatively coupled two-mode magnon system. Two YIG (Yttrium Iron Garnet) spheres are parametriclly pumped by two cavities modes. Each magnon only couples to one cavity mode and is on the node of the other cavity mode.  The loss rates of cavity  and  magnon are denoted by $\gamma_{\rm c}$ and $\gamma_{\rm m}$, respectively. The blue arrow represents the light field driven by the medium, with  frequency  $\omega_{\rm d}$  and  amplitude  $G$. Therefore, two magnons are independently pumped. A third lossy cavity mode couples to both magnons and induces collective loss.}
\label{two_mode system}
\end{figure}

\begin{figure}[t]
\centering
\includegraphics[width=3in]{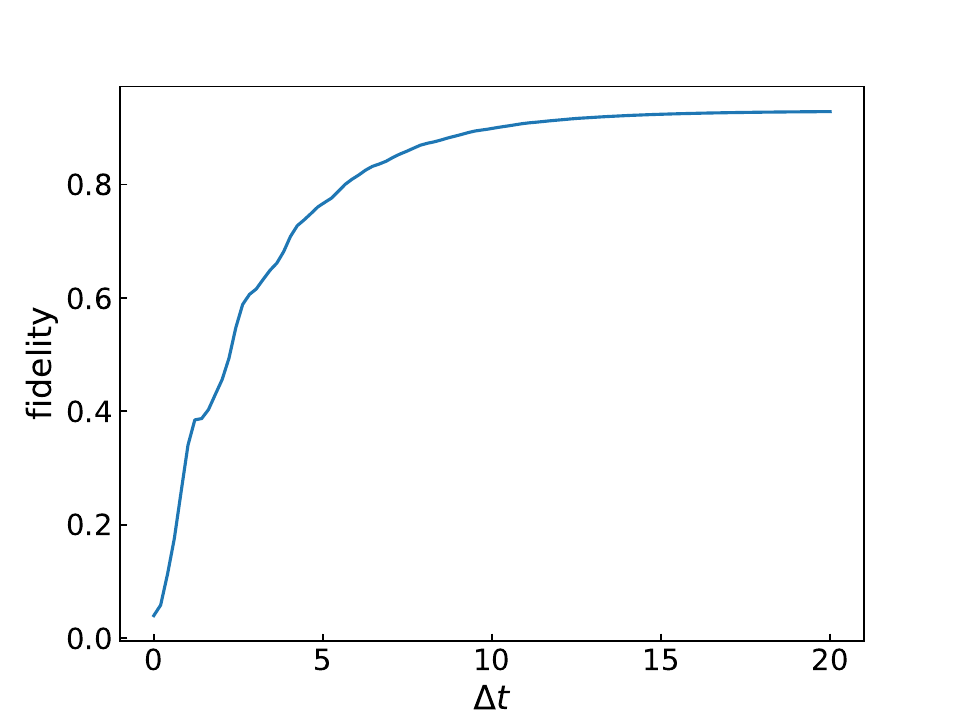}
\caption{The fidelity with respect to an ideal entangled cat state of the two-magnon system with collective loss. The parameters are set to be $\Delta =1$, $S=1.8$, $K=1.2$, $g=0$, and $\gamma_{\rm c}=5$. Note the the parameters are given in unit of $\Delta$. The Fock space truncation is $N=15$. It can be seen that the fidelity of the system gradually increases in time towards a maximum at $0.93$ .}
\label{fidelity_inside}
\end{figure}

In Sec.~\ref{Cat state}, we showed that a single-magnon system with a two-photon pump and Kerr nonlinearity can exhibit strong non-Gaussian properties. Therefore, it is natural to expect the possibility of non-Gaussian entanglement in such systems, too. Following this idea, we consider a two-magnon system with dissipative coupling, as illustrated in Fig.~\ref{two_mode system}. The uncoupled system consists of two YIG  spheres in two parametrically driven microwave cavities (alternatively, two modes in a single cavity). We obtain an effective Hamiltonian composition similar to that of the single-mode magnon system after adiabatically eliminating the Hamiltonian of the two-mode magnon system: 
\begin{eqnarray}\label{two-modeH}
        H&=&S(\hat{b}_{1}^{2} +\hat{b}_{1}^{\dagger2} +\hat{b}_{2} ^{2} +\hat{b}_{2} ^{\dagger2}  )+\frac{K}{2}(\hat{b}_{1} ^{\dagger}\hat{b}_{1} ^{\dagger}\hat{b}_{1}\hat{b}_{1}+\hat{b}_{2} ^{\dagger}\hat{b}_{2} ^{\dagger}\hat{b}_{2} \hat{b}_{2})\nonumber\\
        &&+\Delta(\hat{b}_{1}^{\dagger}\hat{b}_{1} +\hat{b}_{2}^{\dagger} \hat{b}_{2})
        +g(\hat{b}_{1}^{\dagger}\hat{b}_{2} +\hat{b}_{1} \hat{b}_{2}^{\dagger}).
\end{eqnarray}
Here the detuning and the effective nonlinear pump intensity follow the same definition as the single-mode system in  Sec.~\ref{Cat state}, i.e., $\Delta = \Delta _{\rm m}-\left |g\right | ^{2}\Delta _{\rm c}/ (\Delta _{\rm c}^{2}-\left |G \right | ^{2})+\Delta_{\rm other} $ and $S = \left | g\right | ^{2} G / 2(\Delta _{\rm c}^{2}-\left |G\right | ^{2}) $. The coupling term describes the undesired cross-talk introduced by the coupling channel or other cavity modes (see also Appendix \ref{Dissipative channel}).  Note that the phase of the two-photon pump is different from the one in Eq.~(\ref{singlenonlinearmagnon}). A third cavity (or cavity mode), which couples to both magnon modes, can provide the coupling required to generate the entanglement. We introduce detuning and loss to the third cavity mode to obtain a dissipative coupling between the two magnon modes (see also Appendix \ref{Dissipative channel}):
\begin{eqnarray}\label{losschannel}
\hat{L}_{\rm c}&=&\sqrt{\gamma_{\rm c}}(\hat{b}_{1}+\hat{b}_{2}),
\end{eqnarray}
with
\begin{eqnarray}\label{dissipativecoupling}
\frac{\mathrm{d} \rho }{\mathrm{d} t} &=&-i[H,\rho ]+\frac{1}{2} (2\hat{L}_{\rm c}\rho\hat{L} _{\rm c}^{\dagger }-\rho \hat{L}_{\rm c} ^{\dagger }\hat{L}_{\rm c}-\hat{L}_{\rm c} ^{\dagger }\hat{L}_{\rm c} \rho ).
\end{eqnarray}
Here, $\gamma_{\rm c}$ is the collective loss rate.
Such a dissipative coupling can generate non-Gaussian entanglement based on the presence of non-Gaussian single-mode states in  single magnons. According to the findings in Fig.~\ref{bestlastzzy}, the Hamiltonian~(\ref{two-modeH}) can give rise to the following separable two-mode cat state,
\begin{equation}
|\psi\rangle_{\rm sep}=\frac{1}{2+\epsilon_{\rm sep}}\left ( \left | \alpha   \right \rangle+\left | -\alpha   \right \rangle   \right ) \otimes \left ( \left | \alpha   \right \rangle +\left |-\alpha  \right \rangle  \right ),\label{two mode cat state}
\end{equation}

where the correction $\epsilon_{\rm sep}$ is due to the overlap between the coherent states, $\langle -\alpha|\alpha\rangle\neq0$. It is straightforward to evaluate the effect of the collective loss channel~(\ref{dissipativecoupling}) on each term in Eq.~(\ref{two mode cat state}),
\begin{equation}
   \begin{split}
&\hat{L}_{\rm c}\left ( \left | \alpha  \right \rangle  \otimes \left | \alpha   \right \rangle  \right ) =2\sqrt{\gamma_{\rm c} }\alpha \left ( \left | \alpha  \right \rangle\otimes \left | \alpha  \right \rangle \right),\\
&\hat{L}_{\rm c} \left ( \left | \alpha  \right \rangle  \otimes \left | -\alpha   \right \rangle  \right )=0,\\
&\hat{L}_{\rm c}\left ( \left | -\alpha  \right \rangle  \otimes \left | \alpha   \right \rangle  \right )=0,\\
&\hat{L}_{\rm c} \left ( \left | -\alpha  \right \rangle  \otimes \left | -\alpha   \right \rangle  \right )=2\sqrt{\gamma_{\rm c} }\alpha \left ( \left | -\alpha  \right \rangle\otimes \left | -\alpha  \right \rangle \right ).\\
\end{split}
\end{equation}
We find that the Lindblad operator $\sqrt{\gamma_{\rm c}}(\hat{b}_{1}+\hat{b}_{2})$ can suppress the $|\alpha\rangle\otimes|\alpha\rangle$ term and the $|-\alpha\rangle\otimes|-\alpha\rangle$ term in the separable state in Eq.~(\ref{two mode cat state}), leaving the other two terms as a dark state. As a result, we can expect the generated state to be
\begin{equation}\label{twoment}
|\psi\rangle_{\rm ent}=\frac{1}{\sqrt{2+\epsilon_{\rm ent}}} \left (|\alpha \rangle \otimes| -\alpha\rangle+|-\alpha  \rangle \otimes|\alpha \rangle \right ),
\end{equation}
which is a paradigmatic non-Gaussian entangled state. To confirm this expectation, we simulate the evolution of the system and evaluate the fidelity with the state (\ref{twoment}) in Fig.~\ref{fidelity_inside}. 

In our numerical simulations, we set $\Delta=1$, $S=1.8$, $K=1.2$, $\gamma_{\rm c}=5$, $g=0$, and assume a vacuum initial state. Note that, we set $\alpha$ in the target state (\ref{twoment}), to $1.4i$. Inspecting Fig.~\ref{fidelity_inside}, we find that the fidelity of the two-mode magnon system with respect to our target state (\ref{twoment}) gradually increases in time towards the maximum value $0.93$. Note that with the dissipative coupling, the entangled cat state is a steady state of the system, while the cat states in the single-mode case are transient  due to the unitary evolution. Furthermore, we stress that the entanglement here is between two magnon modes instead of  between a magnon mode and a photon mode~\cite{magnon8}.

\subsection{Non-Gaussian entanglement criterion}
After demonstrating the possibility of non-Gaussian entanglement in our model with fidelities as in Fig.~\ref{fidelity_inside}, we now aim at a rigorous criterion on whether the two-mode magnon system is in an entangled state. Reliable methods are entanglement witnesses{\color{red},} specifically, in our case, Bell inequalities~\cite{inequality1,inequality2}. Consider a two-qubit system, then the entanglement between the two qubits is vertified if the entanglement qualifier,
\begin{equation}\label{bell1}
   \begin{split}
\mathcal{Q}=\left \langle\hat{A} _{\rm 0}\otimes\hat{B} _{\rm 0}\right \rangle + \left \langle\hat{A} _{\rm 0}\otimes\hat{B} _{\rm 1}\right \rangle - \left \langle\hat{A} _{\rm 1}\otimes\hat{B} _{\rm 0}\right \rangle + \left \langle\hat{A} _{\rm 1}\otimes\hat{B} _{\rm 1}\right \rangle,
\end{split}
\end{equation}
exceeds the threshold value $2$. Here the the local observables are defined as
\begin{eqnarray}
  \hat{A} _{\rm 0}=\sigma _{\rm z},\hat{A}_{\rm 1}=\sigma _{\rm x},\hat{B}_{\rm 0}=-\frac{\sigma_{\rm x}+\sigma _{\rm z}}{\sqrt{2}},\hat{B}_{\rm 1}=\frac{\sigma_{\rm x}-\sigma _{\rm z}}{\sqrt{2}},\nonumber
\end{eqnarray}
with $\langle O\rangle$ refering to the expectation value of the observable $O$. Note that the Bell inequality in Eq.~(\ref{bell1}) is designed to detect specific forms of entanglement, and is maximally violated for the following state,
\begin{equation}
   \begin{split}
   \left | \psi \right \rangle = \frac{\left | 0  \right \rangle\otimes \left |1 \right \rangle +\left | 1  \right \rangle \otimes \left | 0  \right \rangle  }{\sqrt{2} },
   \end{split}
\end{equation}
which has a similar form to the state~(\ref{twoment}) in our coupled two-mode system. In spite of the similarity of these states, the Bell inequality cannot be directly applied to our system, because of the differing Hilbert spaces. To resolve this obstacle, we utilize modular variables~\cite{modular1,modular2,modular3,modular4,modular5}, which allow us to map cat-like states to qubit states. To this end, note that the entangled cat states~(\ref{twoment}) consist of several coherent states $|\pm\alpha\rangle$, which are the straightforwardly captured by modular variables, featuring only a single ``integer component''~\cite{modularvariable1,modularvariable2} in the large photon number limit $|\alpha\gg 1|$.

\begin{figure}[t]
\centering
\includegraphics[width=0.5\textwidth]{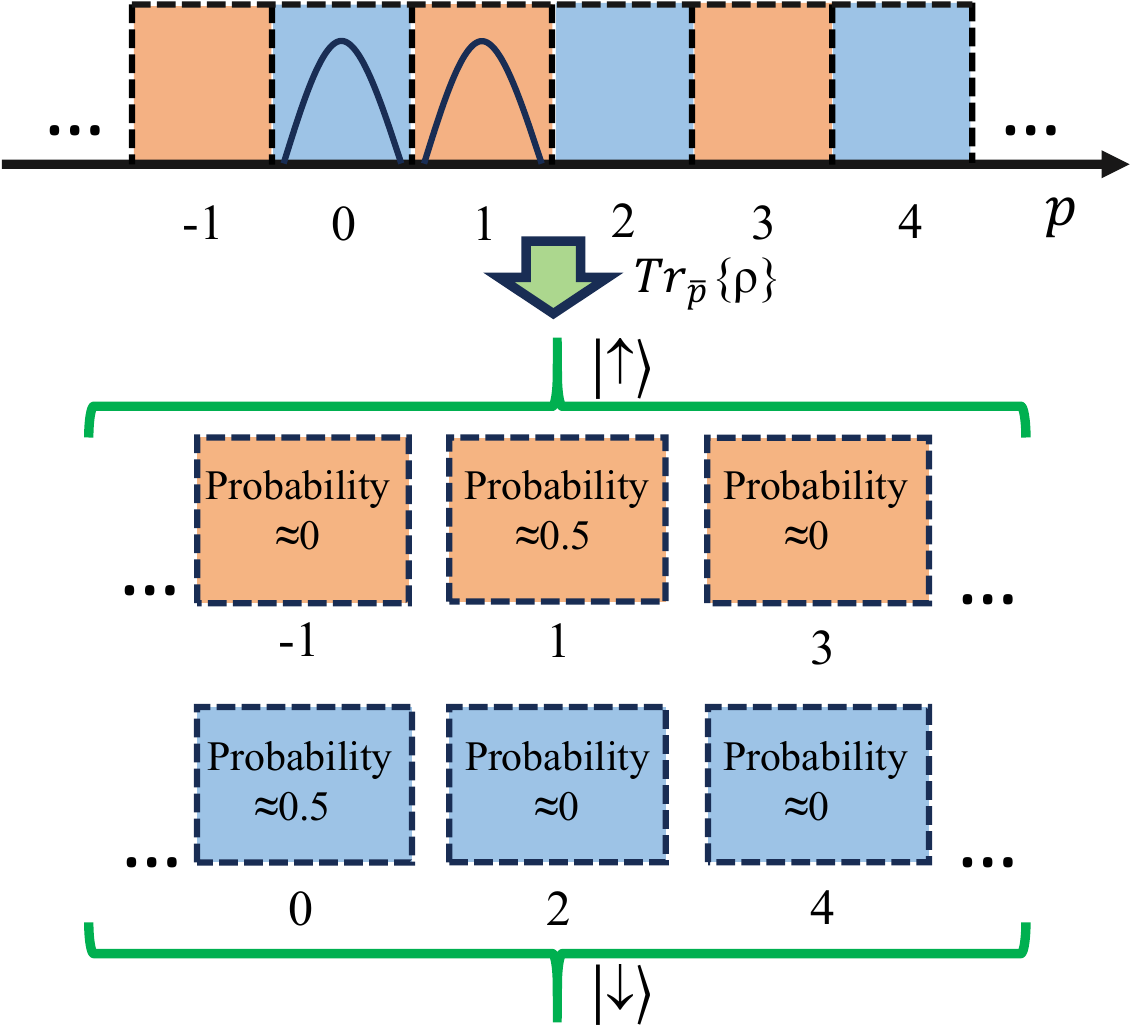}
\caption{Illustration of modular-variable-based projection. The continuous variable is divided into an integer part (dashed box) and a modular part (the wave function component inside a specific dashed box). We trace out the modular part  to obtain a discrete system. The even integer indices and the odd integer indices are grouped together to form two spin states. Note that a suitable choice of the modular length (grid spacing) assigns the two components of a cat-like state to two distinct boxes.}
\label{function}
\end{figure}

The basic idea of modular variables is illustrated in Fig.~\ref{function}. Note that the entangled cat state generated in the previous section is a superposition of two components in the relative momentum space $(p_{\rm r}\equiv p_{1}-p_{2})$. In addition, the reduced states in each subspace also consist of two components in the momentum space, as illustrated in the upper part of Fig.~\ref{function}. To capture the information encoded in such states, we first express the state in the eigenbasis of the dimensionless momentum operator $\hat{p}$ in Eq.~(\ref{ndxp}). The momentum space is then sub-divided into a grid, so that a momentum eigenvalue is decomposed into the grid index $N_{\rm p}$ and the position relative to the grid boundary $\bar{p}$,
\begin{equation}\label{mvdproduct}
   \begin{split}
   \left | \bar{p}   \right \rangle \otimes \left | N_{\rm p}   \right \rangle \equiv \left | p=\bar{p}+l_{\rm p}N_{\rm p}   \right \rangle.
   \end{split}
\end{equation}
The grid number $N_{\rm p}$ is called the \emph{integer momentum} and $\bar{p}$ is called the \emph{modular momentum}. By introducing the modular variables, the total space may now be thought of as the direct product of two subspaces. Note that the direct product form in Eq.~(\ref{mvdproduct}) is different from the rotor picture commonly used in the modular variable approach~\cite{modular1,modular2,modular3,modular4,modular5}, which is formed by the direct product of modular position and modular momentum $|\bar{p}\rangle\otimes|\bar{x}\rangle$. To capture the qubit properties of a cat state, it is necessary to choose an appropriate grid spacing $l_{\rm p}$, as illustrated in Fig.~\ref{function}. Specifically, the optimal choice for the state in Eq.~(\ref{twoment}) is decided by the amplitude

\begin{equation}
   \begin{split}
l_{\rm p}^{\rm opt} =2\sqrt{2}|\alpha|.
\end{split}
\end{equation}

This follows from the fact that a reduced single-mode coherent component $|\alpha\rangle$ in Eq.~(\ref{twoment}) is a Gaussian distribution centered at $(p=\sqrt{2}|\alpha|)$ in the momentum space. Therefore, the separation between the centers of these two components, i.e., $|\alpha\rangle$ and $|-\alpha\rangle$, is $2\sqrt{2}|\alpha|$. Due to the translational invariance of the modular variable~\cite{modularvariable1,modularvariable2}, the optimal modular length given by this value renders the modular parts identical for all the grid segments.
Through the grid decomposition, the qubit-like properties of a cat-like state are transferred to the integer momentum subspace, while the `shape of each qubit state' is preserved in the modular momentum subspace. As our goal is map a continuous variable space to a qubit space, we first trace out the modular momentum,
\begin{equation}
   \begin{split}
\rho _{\rm discrete} =\int_{0}^{l_{\rm p} }d\bar{p}   \left \langle\bar{p}| \rho  |\bar{p} \right \rangle .
\end{split}
\end{equation}
The space after tracing is discrete  infinite-dimensional. To obtain the desired qubit space, we further group this discrete space into the even grid indeces and the odd grid indeces, as illustrated in Fig.~\ref{function}. By doing so, the space can be described by the direct product of a cell space denoted by an integer $m$ and an effective spin space denoted by $n=0,1$,
\begin{equation}
   \begin{split}
\left | m   \right \rangle _{\rm cell} \otimes \left | n   \right \rangle _{\rm es} \equiv \left |N_{\rm p} =2m+n+1   \right \rangle.
\end{split}
\end{equation}
This discrete space can be mapped to an effective spin space by tracing out the cell space~\cite{modular6}:
\begin{eqnarray}\label{esprojection}
\rho _{\rm es} &=& \sum_{m}^{}   \left \langle m| _{\rm cell} \rho_{\rm discrete}   |m \right \rangle _{\rm cell} \nonumber\\&=&\sum_{m}^{} \int_{0}^{l_{\rm p} } d\bar{p} \left\langle m\right| _{\rm cell} \otimes \left \langle \bar{p}| \rho  | \bar{p}   \right \rangle\otimes \left | m \right \rangle_{\rm cell}.
\end{eqnarray}
After mapping the Hilbert spaces of both magnon modes to spin spaces, the Bell inequality can be applied to detect the entanglement. While direct measurement of the effective spin state $\rho_{\rm es}$ may be experimentally challenging, it can be inferred from the original continuous-variable state $\rho$, which can be reconstructed through quantum  state tomography, according to Eq.~(\ref{esprojection}).

Note that a simpler mapping, based on a division into a positive and a negative subspace, could also be applied here,
\begin{eqnarray}
   \left | |p|   \right \rangle \otimes \left | {\rm sign}(p)   \right \rangle &\equiv& \left | p= {\rm sign}(p)|p|    \right \rangle,\nonumber\\
   \rho _{\rm es} &=&\int_0^{\infty}d|p|\langle |p||\rho||p|\rangle.
\end{eqnarray}
However, the modular-variable-based projection can also deal with states with more complicated structures, like Gottesman-Kitaev-Preskill states~\cite{10.1103/PhysRevLett.125.040501}.

Note that the Bell inequality considered here only represents a specific case of a more general family of Bell inequalities~\cite{bellpaper1,bellpaper2,bellpaper3,bellpaper4}, which can take several different forms (see also Appendix ~\ref{Bell state measurement}).
\begin{figure}[t]
\centering
\includegraphics[width=3in]{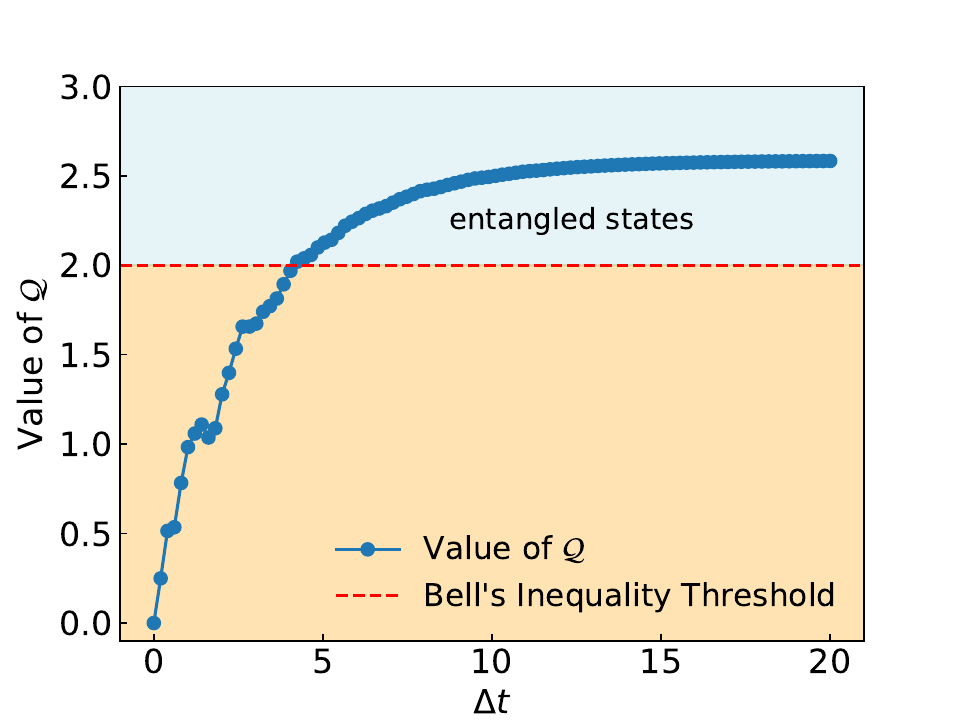}
\caption{ Time evolution of the entanglement qualifier of the two-magnon system after the modular variable-based projection. The parameters are set to be $S=1.8$, $K=1.2$, $g=0$, and $\gamma_{\rm c}=5$. Note the the parameters are provided in units of $\Delta$. The Fock space truncation is $N=15$.  Values above the entanglement threshold  $\mathcal{Q}>2$ red dashed line indicate that the entanglement encoded in the continuous variable space is successfully converted to the effective spin space.}
\label{judge_small}
\end{figure}
\section{Numerical simulations}\label{2}
\subsection{Verifying the entanglement preservation in the effective spin subspace}
We first verify that the projection based on the modular variables can preserve the entanglement present in the original space. To show this, we perform numerical simulations with the parameters used in Fig.~\ref{fidelity_inside}, and apply  Bell inequality to the state obtained after the projection. 

According to the numerical results in Fig.~\ref{judge_small}, the entanglement of the original state can be detected by the Bell inequality after the projection as the value of the qualifier $\mathcal{Q}$ can surpass the entanglement threshold $2$. It is not surprising  that the evolution of the entanglement qualifier is similar to the one of the fidelity with respect to an ideal entangled state in Fig.~\ref{fidelity_inside}. However, we emphasize that the Bell inequality cannot be directly applied to an entangled cat state. The asymptotic value for the qualifier is $Q\approx2.58$, which is slightly below  the saturation value of  Bell inequality at 2.82. There are two possible reasons for this reduction. Firstly the generated entangled cat state  is not  perfect.  Secondly, there is a nonvanishing overlap between two coherent states with opposite phases $|\alpha\rangle$ and  $|-\alpha\rangle$.

\begin{figure}[t]
\centering
\includegraphics[width=3in]{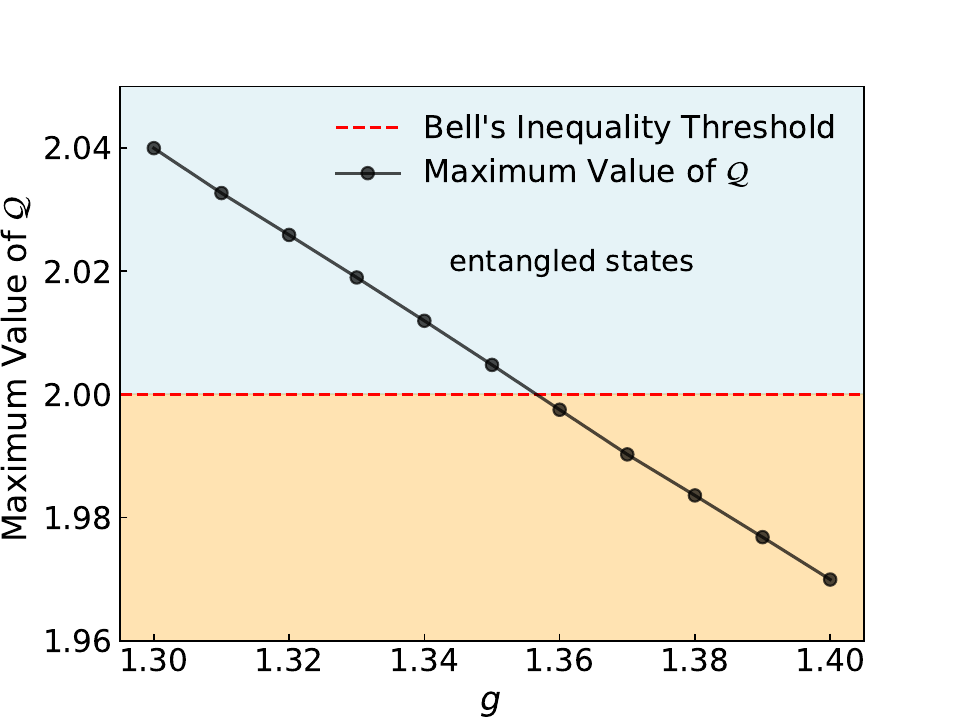}
\caption{ Relation between the maximum value of the  entanglement qualifier $\mathcal{Q}$, which is achieved by the two-magnon system, and the cross-talk coupling term. The threshold value of the qualifier is marked with the red dashed line. The parameters are set to be $S=1.8$, $K=1.2$, $\gamma_{\rm s}=0$, and $\gamma_{\rm c}=5$. Note that the parameters are given in unit of $\Delta$. The Fock space truncation is $N=15$. The results indicate a comparatively high tolerance to the cross-talk coupling.}
\label{g_inside}
\end{figure}

\subsection{The influence of the crosstalk between the magnon modes}
One potential problem for the two YIG  spheres in a common cavity can be the cross talk caused by the cavity modes or direct coupling. To address this, we study the influence of the coupling between two magnon modes on the entanglement generation by considering a nonvanishing $g$ in the Hamiltonian~(\ref{two-modeH}). For comparison with the results in the previous sections, the parameters are set as in Figs.~\ref{fidelity_inside} and \ref{judge_small}. We simulate the dynamical evolution with different coupling strengths $g$ and pick up the maximum values of the entanglement qualifier $Q$ achieved during the evolution. If these maximum values are larger than $2$, entanglement can be generated with  the corresponding set of parameters as Bell inequality is a sufficient condition for entanglement. The results in Fig.~\ref{g_inside} indicate that entanglement can be detected in the system for $g<1.35$. Note that such a cross-talk coupling strength limit is remarkably high, as the maximum tolerable coupling strength is $ g_{max}\approx1/3\gamma_{\rm c} $. Therefore, we can conclude that the cross-talk does not represent a critical issue in generating entanglement in most cases, as the tolerable strength is comparable to other couplings in the system.

\begin{figure}
\centering
\includegraphics[width=0.5\textwidth]{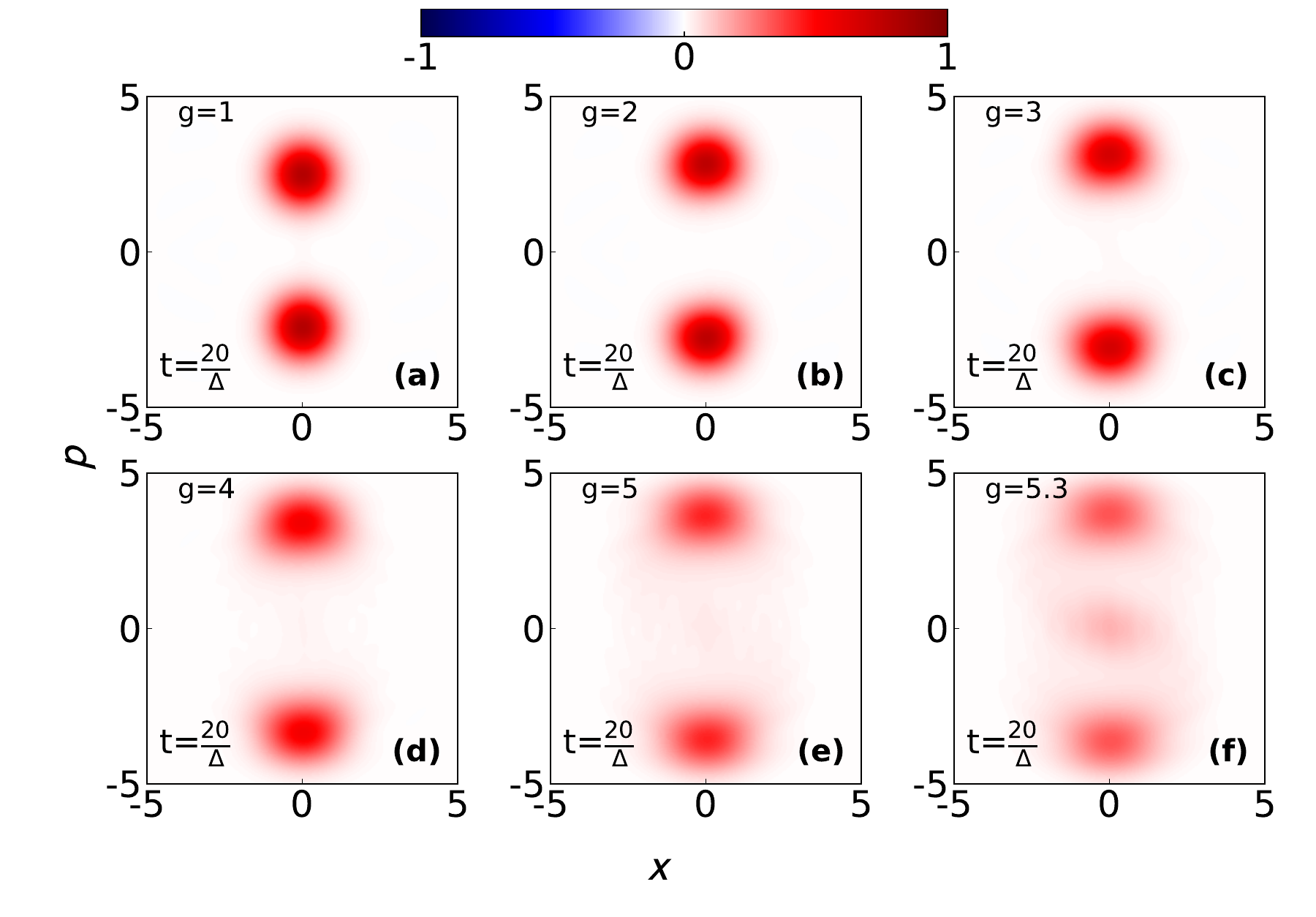}
\caption{ Wigner function of the reduced system state obtained by tracing out one magnon mode. The results are calculated with different cross-talk coupling strengths at the time $t$ = $20/\Delta$. The plots demonstrate how the steady state is disturbed by the coupling.}
  \label{fig:whole}
\end{figure}
To have a more intuitive understanding of the cross-talk coupling, we trace out one magnon mode and evaluate the Wigner function of the reduced state in Fig.~\ref{fig:whole}. Note that the two magnon modes are symmetric in our study, and there is no interference pattern after tracing. Inspecting Fig.~\ref{fig:whole}, we  find that strong couplings disturb the steady state of the system. The results in Figs.~\ref{fig:whole}(d), (e), and (f) deviate significantly from the coherent components in typical cat states. For the weaker coupling strengths featured in Figs.~\ref{fig:whole}(a), (b), and (c), there is no visible change. We conclude that the influence of the cross talk is  mainly  on the coherence as the entanglement disappears at $g\approx1.35$. 
\subsection{\label{sec:level2}Influence of  single-photon loss}
\begin{figure}[t]
\centering
\includegraphics[width=3in]{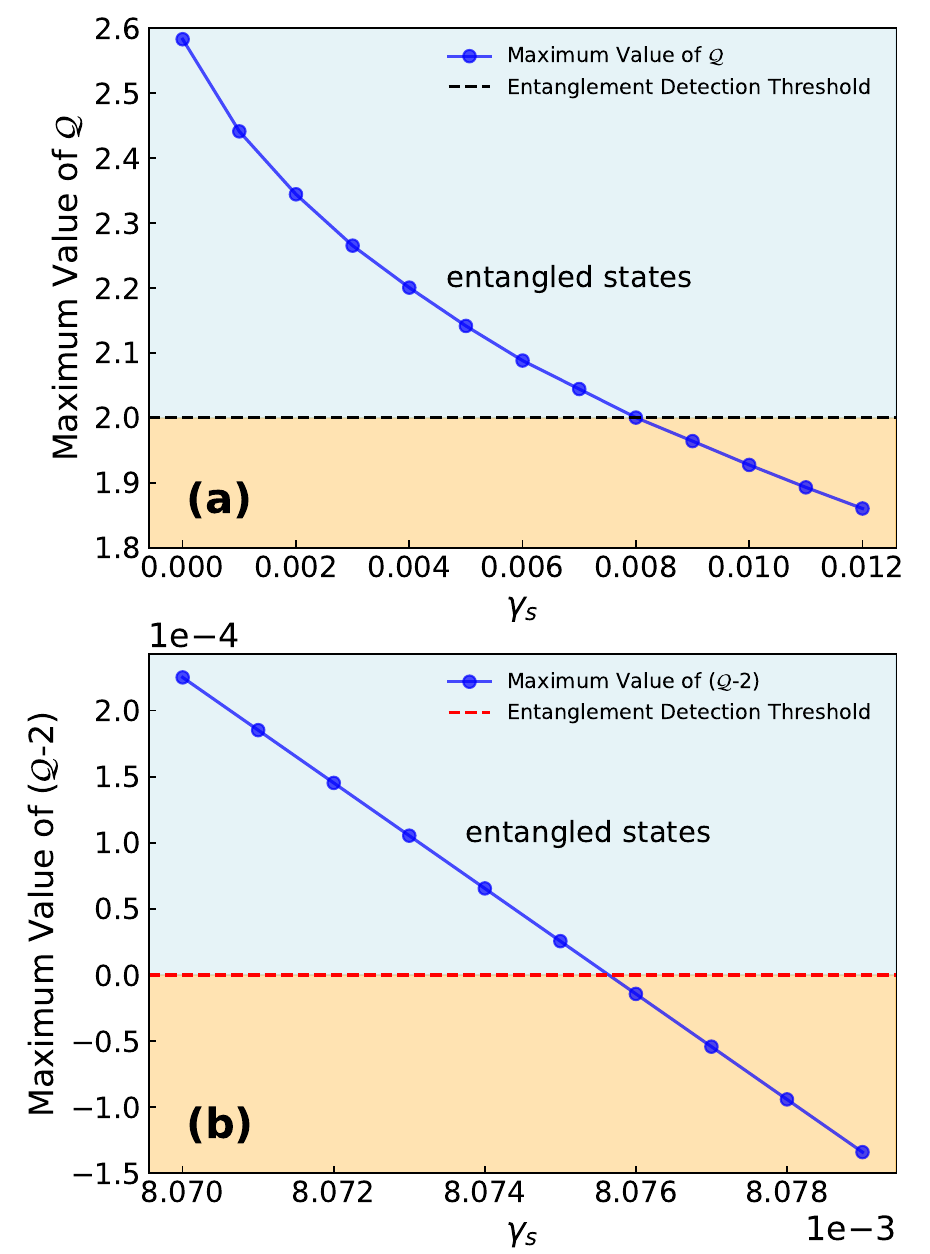}
\caption{ Relation between the highest achievable values of the entanglement qualifier and the single-photon loss rate. The parameters are set to be $S=1.8$, $K=1.2$, $g=0$, and $\gamma_{\rm c}=5$. Note the the parameters are given in units of $\Delta$. The Fock space truncation is $N=15$. (a) The single-photon loss rate is varied over a wide range, and  the maximum values of the qualifier are shown. (b) A finer scan of single-photon loss rates in the vicinity of the entanglement threshold $\mathcal{Q}=2$ and the qualifier  presented as $\mathcal{Q}-2$. The results indicate an upper limit for the single-photon loss rate of about $0.008$ and a linear dependence of the qualifier on $\gamma_{\rm s}$ near the threshold.}
\label{gamma_inside}
\end{figure}

Single-photon loss is usually the main obstacle for generating entanglement~\cite{single-photon1,single-photon2,single-photon3,single-photon4,single-photon5}. Therefore, we now study the maximum tolerable single-photon loss for generating entanglement in the two-mode magnon system.  Single-photon loss is described by the following Lindblad terms,
\begin{eqnarray}
L_{{\rm s},i}=\hat{a}_i,
\end{eqnarray}
with the single-photon loss rate $\gamma_{\rm s}$. Recall that the contribution of the Lindblad operator is described by the master equation~(\ref{dissipativecoupling}). Under the influence of the single-photon loss,  entanglement is generically destroyed in the long-time limit. To access this, we take again the maximum values of the entanglement qualifier during the dynamical evolution as indicators for successful entanglement generation. The maximum values for the entanglement qualifier $\mathcal{Q}$ are shown in Figs.~\ref{gamma_inside}(a) and (b). Inspecting Fig.~\ref{gamma_inside}(a), we find that the maximum achievable values of the qualifier monotonically decrease with the single-photon loss rate $\gamma_{\rm s}$ within the parameter range studied. In addition, an approximate range $\gamma_{\rm s}\leq0.008$ for successful entanglement generation in the two-mode magnon system can be inferred. A finer parameter scan around the entanglement threshold $\mathcal{Q}=2$, shown in Fig.~\ref{gamma_inside}(b), provides a more precise threshold value for the single-photon loss. Note that the finer parameter scan confirms the monotonic influence of the single-photon loss. As states that violate Bell inequality form a strict subset of all entangled states, the maximum tolerable single-photon loss here is significantly smaller than the one obtained for a modular-variable-base entanglement criterion~\cite{10.1103/PhysRevA.104.013715}.     

In this section, we studied the influence of the cross-talk and the single-photon loss on the capacity of a two-magnon system to generate entanglement. From the results in Figs.~\ref{g_inside} and \ref{fig:whole}, we can conclude that the entanglement is not sensitive to the cross-talk coupling strength $g$. This may follow from the stabilizing effects of the collective loss in Eq.~(\ref{dissipativecoupling}). However, the entanglement is vulnerable to the single-photon loss in each mode. From Figs.~\ref{gamma_inside}(a) and ~\ref{gamma_inside}(b), it can be seen that already a modest single-photon loss rate can inhibit the Bell-inequality violation.
\section{Conclusions}
\label{conclusion}\label{3}
 We studied the generation of non-Gaussian entanglement in magnon systems.  The target non-Gaussian state is a cat-like state, which can be generated by nonlinear pumping above the threshold, that is, in the parametric unstable regime. We first showed that a transient single-mode cat state can be generated under the influence of the nonlinear pump and the Kerr-nonlinearity. Next, we considered the two-mode case with dissipative coupling. Our numerical results indicate that a steady entangled cat state can be generated with the dissipative coupling. To access the entanglement in the state, we projected the system state to an effective spin space and applied a Bell inequality. Finally, we studied the impact of detrimental effects like cross-talk coupling or single-photon loss, and provided the necessary parameters for violating Bell inequality.

Our work addresses the entanglement generation in magnon systems in a less explored regime, i.e., the parametric unstable regime. We believe that this  can contribute to the study of entanglement in hybrid systems, which represents an important potential application for magnon systems, by mapping the cat-like entanglement to the entanglement in qubit systems.
\section{Acknowledgments}
This research was supported by Zhejiang Provincial Natural Science Foundation of China under Grant No.~LQ24A050002; Foundation of Department of Scicence and Technology of Zhejiang Province, No. 2022R52047; the National Natural Science Foundation of China (NSFC) (Grant No.~12075209); the Innovation Program for Quantum Science and Technology (2023ZD0300904). C.G. is partially supported by a RIKEN Incentive Research Grant.
\appendix
\section{Parametric unstable regime}
Consider the following parametrically pumped system
\begin{eqnarray}
H=\Delta\hat{a}^{\dag}\hat{a}+S({\hat{a^\dag}}^2+\hat{a}^2),
\end{eqnarray}
with $\Delta>0$. For negative detuning $\Delta<0$, we can consider the Hamiltonian $H'=-H$. Such a Hamiltonian can be diagonalized as
\begin{eqnarray}
H=(c_1\hat{a}^{\dag}+c_2\hat{a})(c_1\hat{a}^{\dag}+c_2\hat{a})^{\dag},
\end{eqnarray}
with,
\begin{eqnarray}
c_1^2+c_2^2&=&\Delta,\nonumber\\
c_1c_2&=&S.
\end{eqnarray}
The effective mode $c_1\hat{a}^{\dag}+c_2\hat{a}$ is called squeezed mode. However, the validity of such modes requires,
\begin{eqnarray}
\Delta>2S.
\end{eqnarray}
When this relation is not satisfied, the system enters the parametric unstable regime.
\section{The effective nonlinear pump of the magnon}\label{adiabatic elimination}
In this appendix we present the adiabatic elimination  for single-mode magnon subsystems. Consider a magnon in a parametrically pumped cavity:
\begin{equation}
   \begin{split}
   &H=\Delta _{\rm c}\hat{a}^{\dagger}\hat{a} +\Delta _{\rm m}\hat{b}^{\dagger}\hat{b}+\frac{K}{2}\hat{b}^{\dagger}\hat{b}^{\dagger}\hat{b}\hat{b}+g(\hat{a}^{\dagger} \hat{b}+\hat{a}\hat{b}^{\dagger})+\\&\frac{G}{2}(\hat{a}^{\dagger}\hat{a}^{\dagger}+\hat{a}\hat{a}), \\
   \end{split}
\end{equation}
which satisfies: $\Delta _{\rm c(m)}=\omega_{\rm c(m)} -\frac{\omega_{\rm d} }{2} $, and based on the system Hamiltonian quantities we can obtain the Heisenberg equations for $\hat{a}$ and $\hat{b}$ respectively:
\begin{equation}
   \begin{split}
   &\dot{\hat{a} }=-i\Delta _{\rm c}\hat{a}-ig\hat{b}-iG\hat{a}^{\dagger},\\
   &\dot{\hat{b} }=-i\Delta _{\rm m}\hat{b}-iK\hat{b}^{\dagger}\hat{b}\hat{b}-ig\hat{a}. \\
\end{split}                                                 
\end{equation}
When the detuning of the cavity is large{\color{red},} $\Delta_{\rm c}\gg g$, we can assume $\dot{\hat{a}}$= 0 and $\dot{\hat{a}} ^{\dagger}$ = 0 to obtain:
\begin{equation}
   \begin{split}
   &\hat{a}^{\dagger} =\frac{gG\hat{b}-\Delta _{\rm c} g\hat{b}^{\dagger}}{\Delta _{\rm c}^{2}-\left |G \right | ^{2}    }, \\
   &\hat{a}  =\frac{gG\hat{b}^{\dagger}-\Delta _{\rm c} g\hat{b}}{\Delta _{\rm c}^{2}-\left |G \right |^{2}}, \\
    \end{split}
\end{equation}
The equation for  $\hat{b} ^{\dagger}$ then becomes:
\begin{equation}
   \begin{split}
   &\dot{\hat{b} }^{\dagger}=i\Delta _{\rm m}\hat{b}^{\dagger}+iK\hat{b}\hat{b}^{\dagger}\hat{b}^{\dagger} +ig\frac{gG\hat{b}-\Delta _{c} g\hat{b}^{\dagger}}{\Delta _{\rm c}^{2}-\left |G \right |^{2}}.\\
    \end{split}
\end{equation}
Note that we have the relation
\begin{eqnarray}
\dot{\hat{b}}^{\dag}&=&i[H,\hat{b}^{\dag}],
\end{eqnarray}
so that the corresponding effective Hamiltonian for the magnon mode is
\begin{equation}
   \begin{split}
  H=&(\Delta _{\rm m}-\frac{\left |g\right | ^{2}\Delta _{\rm c} }{\Delta _{\rm c}^{2}-\left |G \right | ^{2}   } ) \hat{b}^{\dagger}\hat{b}+\frac{K}{2}\hat{b}^{\dagger}\hat{b}^{\dagger}\hat{b}\hat{b}\\+&\frac{1}{2} \frac{\left |g\right | ^{2}G }{ \Delta _{\rm c}^{2}-\left |G \right | ^{2}   }(\hat{b}^{\dagger}\hat{b}^{\dagger}-\hat{b}\hat{b}). \\\label{e}    
    \end{split}
\end{equation}
\section{Dissipative coupling channel and cross-talk coupling}\label{Dissipative channel}
In the study of two-mode magnon systems, we  introduce a dissipative channel to obtain our desired target state. In the following we present the   derivation of the dissipative channel in Eq.~(\ref{losschannel}).

Consider the coupling term between a lossy cavity mode $\hat{a}$ and the two magnon modes $\hat{b}_{1}$ and $\hat{b}_{2}$,
\begin{equation}
   \begin{split}
   g(\hat{b}_{1}+\hat{b}_{2}) \hat{a}  ^{\dagger} +g\hat{a} (\hat{b}_{1}^{\dagger}+\hat{b}_{2}^{\dagger}),
\end{split}
\end{equation}
The Langevin equations for the magnon modes and the cavity are:
\begin{equation}
   \begin{aligned}
  \dot{\hat{b}}_{1} &= i[H,\hat{b}_{1} ]= -ig\hat{c}, \\
   \dot{\hat{b}}_{2} &=i[H,\hat{b}_{2} ]= -ig\hat{c}, \\
   \dot{\hat{a}}\;&=i[H,\hat{a} ]-\gamma_{\rm c} \hat{a} +\sqrt{2\gamma_{\rm c} } C_{\rm in} \\
   &= -ig(\hat{b}_{1} +\hat{b}_{2})-\gamma_{\rm c}\hat{a}+\sqrt{2\gamma_{\rm c} } C_{\rm in}.\\
    \end{aligned}
\end{equation}
If we assume that the loss is strong enough to keep the cavity in the steady state,
\begin{eqnarray}
\hat{a}=\frac{ig(\hat{b}_{1}+\hat{b}_{2}) +\sqrt{2\gamma_{\rm c} } C_{\rm in}}{\gamma_{\rm c}},
\end{eqnarray}
the effective evolutions of the magnons become
\begin{equation}
   \begin{split}
&\dot{\hat{b} }_{1}  =-\frac{g^{2} }{\gamma_{\rm c} }(\hat{b}_{1}+\hat{b}_{2})-\frac{igC_{\rm in} }{\sqrt{\gamma } } ,\\
&\dot{\hat{b} }_{2}  =-\frac{g^{2} }{\gamma_{\rm c} }(\hat{b}_{1}+\hat{b}_{2})-\frac{igC_{\rm in} }{\sqrt{\gamma } } ,\\
   \end{split} 
\end{equation}
which is equivalent to a dissipative channel of the form:\\
\begin{equation}
   \begin{split}
\hat{L}=\sqrt{\gamma_{\rm c} } (\hat{b}_{1}+\hat{b}_{2}).
 \end{split} 
\end{equation}
This concludes our proof of the adiabatic elimination process for single-mode cavity magnon systems.

However, other cavity modes can also couple to these two magnons and cause cross-talk coupling between them. Consider the detuned mode $\hat{d}$ with the Hamiltonian,
\begin{equation}
   \begin{split}
   \Delta \hat{d}^{\dag}\hat{d}+g(\hat{b}_{1}+\hat{b}_{2}) \hat{d}  ^{\dagger} +g\hat{d} (\hat{b}_{1}^{\dagger}+\hat{b}_{2}^{\dagger}).
\end{split}
\end{equation}
The Heisenberg equations for the operators are now,
\begin{equation}
   \begin{split}
   \dot{\hat{b}}_{1} &= -ig\hat{d}, \\
   \dot{\hat{b}}_{2} &= -ig\hat{d}, \\
  \dot{\hat{d}}\; &= -ig(\hat{b}_{1} +\hat{b}_{2})-i\Delta\hat{d}, \\
    \end{split}
\end{equation}
where the loss in the cavity mode $\hat{d}$ is not considered. We can assume that the detuning $\Delta$ is large enough to keep the cavity mode in the steady state,
\begin{eqnarray}
\hat{d}=\frac{-g(\hat{b}_{1}+\hat{b}_{2})}{\Delta}.
\end{eqnarray}
The effective equations for the magnons then become
\begin{equation}\label{cross-talk coupling}
   \begin{split}
   \dot{\hat{b}}_{1} &=i\frac{g^2(\hat{b}_{1}+\hat{b}_{2})}{\Delta}, \\
   \dot{\hat{b}}_{2} &= i\frac{g^2(\hat{b}_{1}+\hat{b}_{2})}{\Delta}, \\
    \end{split}
\end{equation}
which contains a cross-talk term $g^2/\Delta (\hat{b}_{1}\hat{b}_{2}^{\dag}+\hat{b}_{1}^{\dag}\hat{b}_{2})$.\\

\section{normalization factor}\label{normalisation factor}
Here, we determine the small correction $\epsilon_{\rm s}$ in the normalization factor in Sec.~\ref{Cat state}:
\begin{eqnarray}
|\psi\rangle=\frac{1}{\sqrt{2+\epsilon_{\rm s}}}(|\alpha\rangle+|-\alpha\rangle).
\end{eqnarray}
The norm of this state is:
\begin{eqnarray}
\left \langle \psi  | \psi  \right \rangle &=&\frac{1}{{2+\epsilon_{\rm s}}}(\left\langle\alpha\right|+\left\langle-\alpha \right|  )(\left | \alpha   \right \rangle +\left |-\alpha   \right \rangle ),\nonumber\\
 &=&\frac{1}{{2+\epsilon_{\rm s}}}(\left \langle \alpha  | \alpha   \right \rangle +\left \langle -\alpha | -\alpha   \right \rangle+\left \langle\alpha  | -\alpha  \right \rangle+\left \langle-\alpha  | \alpha  \right \rangle). \nonumber\\
\end{eqnarray}
Due to the property of  coherent states,
\begin{eqnarray}
\left \langle \beta   | \alpha  \right \rangle =\exp[-\frac{1}{2}(\left | \alpha  \right| ^{2} +\left | \beta   \right |^{2}  )+\beta ^{*}\alpha],
\end{eqnarray}
we have,
\begin{eqnarray}
\left \langle \psi  | \psi  \right \rangle =\frac{1}{{2+\epsilon_{\rm s}}}(2+2e^{-2\left | a \right |^{2}  } )=1. 
\end{eqnarray}
Thus the normalization factor can be obtained in the form:
\begin{eqnarray}
\frac{1}{\sqrt{\left \langle \psi  | \psi  \right \rangle } } =\frac{1}{\sqrt{2+2e^{-2\left | a \right |^{2}  } } }.
\end{eqnarray}
As a result, we obtain $\epsilon_{\rm s}=2e^{-2\left | a \right |^{2}  } $.

The normalization factor for the separable cat state can also be obtained,
\begin{equation}
\begin{split}
|\psi\rangle_{\rm sep}=\frac{1}{2+\epsilon_{\rm sep}}\left ( \left | \alpha   \right \rangle+\left | -\alpha   \right \rangle   \right ) \otimes \left ( \left | \alpha   \right \rangle +\left |-\alpha  \right \rangle  \right ).\label{two mode cat state new}
\end{split}
\end{equation}
Note that the norm can be calculated as,
\begin{equation}
\begin{split}
\left \langle \psi   | \psi   \right \rangle_{\rm sep}&=\left ( \frac{1}{2+\epsilon_{\rm sep}} \right ) ^{2} \left [ \left ( \left | \alpha   \right \rangle +\left |-\alpha  \right \rangle  \right )\otimes \left ( \left | \alpha  \right \rangle +\left | -\alpha   \right \rangle  \right )   \right ]^{\dagger }\cdot \\&\left [ \left ( \left | \alpha   \right \rangle +\left |-\alpha  \right \rangle  \right )\otimes \left ( \left | \alpha  \right \rangle +\left | -\alpha   \right \rangle  \right ) \right ].
\end{split}
\end{equation}
As the norm should be $1$, we have
\begin{equation}
\begin{split}
\left \langle\rm  \psi   | \psi   \right \rangle_{\rm sep} =\left(\frac{2+2e^{-2\left | a \right |^{2}  } }{2+\epsilon_{\rm sep}}\right)^{2}=1,
\end{split}
\end{equation}
Thus the correction is,
\begin{equation}
\begin{split}
\epsilon_{\rm sep}=2e^{-2\left | a \right |^{2}}.
\end{split}
\end{equation}

Finally, we consider the entangled cat state,
\begin{equation}
|\psi\rangle_{\rm ent}=\frac{1}{\sqrt{2+\epsilon_{\rm ent}}} \left (|\alpha \rangle \otimes| -\alpha\rangle+|-\alpha  \rangle \otimes|\alpha \rangle \right ).
\end{equation}
We can obtain its norm as,
\begin{equation}
\begin{split}
 \left \langle \psi  | \psi  \right \rangle _{\rm ent}&=\left ( \frac{1}{\sqrt{2+\epsilon_{\rm ent}}}  \right ) ^{2} \left ( \left |\alpha   \right \rangle \otimes \left | -\alpha  \right \rangle +\left | -\alpha   \right \rangle \otimes \left |\alpha  \right \rangle  \right )^{\dagger }\cdot\\&\left ( \left |\alpha   \right \rangle \otimes \left | -\alpha  \right \rangle +\left | -\alpha   \right \rangle \otimes \left |\alpha  \right \rangle  \right ),
\end{split}
\end{equation}
so that,
\begin{equation}
\begin{split}
 \left \langle \psi  | \psi  \right \rangle _{\rm ent}=\frac{2+2e^{-4\left | \alpha  \right |^{2}  }}{2+\epsilon_{\rm ent}}=1.
\end{split}
\end{equation}
Thus, we get $\epsilon_{\rm ent}=2e^{-4\left | \alpha \right |^{2} }$ .

\section{Bell state measurement}\label{Bell state measurement}
In general, there can be four different Bell states in a pair of two-level systems:
\begin{eqnarray}
\left | \Phi ^{+}  \right \rangle &=&\frac{1}{\sqrt{2} } (\left | 00 \right \rangle +\left |11 \right \rangle ),\nonumber\\
\left | \Phi ^{-}  \right \rangle &=&\frac{1}{\sqrt{2} } (\left | 00 \right \rangle -\left |11 \right \rangle ),\nonumber\\
\left | \Psi  ^{+}  \right \rangle &=&\frac{1}{\sqrt{2} } (\left | 01 \right \rangle +\left |10 \right \rangle ),\nonumber\\
\left | \Psi  ^{-}  \right \rangle &=&\frac{1}{\sqrt{2} } (\left | 01 \right \rangle -\left |10 \right \rangle ).
\end{eqnarray}
To detect the entanglement in these states with Bell inequality:
\begin{equation}
\begin{split}
\left | E\left ( A_{0}B_{0}\right ) +E\left ( A_{1}B_{1}\right ) +E\left ( A_{0}B_{1}\right )-E\left ( A_{1}B_{0}\right ) \right | \le 2 ,
\end{split}
\end{equation}
we need to choose different observables:
\begin{equation}
\begin{split}
\left |\Phi ^{+  }\right \rangle :A_{0}&=\sigma _{\rm z} ,A_{1} =\sigma _{\rm x},B_{0} =\frac{\sigma _{\rm z}-\sigma _{\rm x}}{\sqrt{2} },B_{1}= \frac{\sigma _{\rm x}+\sigma _{\rm z}}{\sqrt{2} },\\
\left |\Phi ^{-}\right \rangle :A_{0}&=\sigma _{\rm x} ,A_{1} =\sigma _{\rm z},B_{0} =-\frac{\sigma _{\rm x}+\sigma _{\rm z}}{\sqrt{2} },B_{1}= \frac{\sigma _{\rm z}-\sigma _{\rm x}}{\sqrt{2} },\\
\left |\Psi  ^{+}\right \rangle :A_{0}&=\sigma _{\rm z} ,A_{1} =\sigma _{\rm x},B_{0} =-\frac{\sigma _{\rm x}+\sigma _{\rm z}}{\sqrt{2} },B_{1}= \frac{\sigma _{\rm x}-\sigma _{\rm z}}{\sqrt{2} } ,\\
\left |\Psi  ^{-}\right \rangle :A_{0}&=\sigma _{\rm x} ,A_{1} =\sigma _{\rm z},B_{0} =\frac{\sigma _{\rm z}-\sigma _{\rm x}}{\sqrt{2} },B_{1}=-\frac{\sigma _{\rm z}+\sigma _{\rm x}}{\sqrt{2} }.\\
\end{split}
\end{equation}
Note that the expectation values are defined as:
\begin{equation}
\begin{split}
E(A_{\rm i},B_{\rm j})=\left\langle\Psi \right| A_{\rm i}\otimes B_{\rm j}\left | \Psi   \right \rangle .
\end{split}
\end{equation}
For each Bell state and the corresponding observables, it is straightforward to check that
\begin{equation}
\begin{split}
E\left ( A_{0}B_{0}\right ) &=\sqrt{2}/2, ~~E\left ( A_{1}B_{0}\right )=-\sqrt{2}/2,\\
E\left ( A_{0}B_{1}\right )&=\sqrt{2}/2,~~E\left ( A_{1}B_{1}\right )=\sqrt{2}/2.
\end{split}
\end{equation}
It results that in an entangled state, Bell inequality can exceed 2 and have a maximum violation value of $2\sqrt{2}$.
\vspace{12pt}
%%%%%%%
%\begin{thebibliography}{99}

%Proposal PST:

%
%\begin{references}
%\bibliographystyle{unsrt}
\bibliography{catstate}
%\end{references}
\end{document}